\documentclass[10pt,a4paper,oneside,english]{elsart}
\usepackage[T1]{fontenc}
\usepackage[latin1]{inputenc}
\usepackage{float}
\usepackage{graphicx}
\usepackage{amssymb}

\makeatletter

\usepackage{babel}
\usepackage{epsfig}
\makeatother
\begin{document}
\begin{frontmatter}

\title{Sector analysis for a FTSE portfolio of stocks}

\author[TCD]{R. Coelho},
\ead{coelhorj@tcd.ie}
\author[TCD]{S. Hutzler},
\author[TCD]{P. Repetowicz}
\author[TCD]{\and P. Richmond}
\address[TCD]{School of Physics, Trinity College Dublin, Dublin 2, Ireland}

\begin{abstract}
Using a portfolio of stocks from the London Stock Exchange FTSE100
index (FTSE), we study both the time dependence of their correlations
and the normalized tree length of the associated minimal spanning
tree (MST). The first four moments of the distribution of correlations
and lengths of the tree are examined in detail and differences in 
behaviour noted. For different economic groups and
industries, clustering is evident.
However comparing the classification used prior to 2006 with that 
introduced in January 2006 it is clear that the new classification, 
apart from one or two notable exceptions, is much more compatible
with the clustering obtained by the MST analysis.
We finally compare the MST for real data with that obtained for a 
synthetic {\it random market}. The latter tree would seem more like
the structure found by Coronnello {\it et al.} for trees based on high
frequency data.
\end{abstract}
\end{frontmatter}

\section{Introduction\label{Sect1}}

During the past decade, many physicists have used techniques of statistical
physics and complexity to study economic and financial problems \cite{Mantegna_Book,Bouchaud_Book}
and the associated networks \cite{Mendes_Book}. Networks play a crucial
role in these systems simply because trading activity generates networks.
Studying stock networks, where the links represent similarities between
stocks, can prove very valuable for portfolio optimization \cite{Mantegna_phy0507006,Onnela_PRE68_2003}.

A challenging problem is the nature of stock time series and, in particular,
the nature of their randomness \cite{Onnela_PRE68_2003,Mantegna_PRE68_046130,Mantegna_EPJB363_2004}.
Recently the theory of random matrices has proved helpful to characterize
the time series \cite{Laloux_PRL_83_1467,Plerou_PRL_83_1471}. In
this paper we use the concept of a minimal spanning tree (MST) proposed
by Mantegna \cite{Mantegna_EPJB193_1999}, to examine the correlations
for stocks from the London FTSE100 index.

We review briefly the method in the next section and explain how we
choose the time parameters in Section \ref{Sect3}. In Section \ref{Sect4},
we use the approach to examine a portfolio of stocks selected from
the London FTSE100 index. In Section \ref{Sect5}, we examine in more
detail the results for individual stock sectors. For different economic groups and
industries, clustering derived in Section \ref{Sect4} is evident. However
comparing in Section \ref{Sect6} the classification used prior to 2006 with that 
introduced in January 2006 it is clear that the new classification, apart
from one or two notable exceptions, is much more compatible with the clustering
obtained by the MST analysis. We finally compare in Section \ref{Sect7} the MST for
real data with that obtained for a synthetic random and close with a few conclusions.

\section{Definitions\label{Sect2}}

Our main goal is to detect any underlying structure of a portfolio,
such as clustering, or identification of key stocks. We start by computing
the correlation coefficient between time series of log-returns of
pairs of stocks. From these correlations we can compute a {\it distance},
for each pair, which is used for the construction of a network with
links between stocks.

The 100 most highly capitalized companies in the UK that comprise
the London FTSE100, represent approximately $80\%$ of the UK market.
From these 100 stocks, we study the time series of the daily closing
price of $N=67$ stocks that have been in the index continuously over
a period of almost $9$ years, starting in $2^{nd}$ August $1996$
until $27^{th}$ June $2005$. This equals $2322$ trading days per
stock. For our analysis of the time dependence of correlations and
distances, time series are divided in small time windows, each with
width $T$, that will overlap each other. The total number of windows
depends on the window step length parameter, $\delta T$.

\subsection{Correlations}

The correlation coefficient, $\rho_{ij}$ between stocks $i$ and
$j$ is given by:

\begin{equation}
\rho_{ij}=\frac{\langle{\bf R}_{i}{\bf R}_{j}\rangle-\langle{\bf R}_{i}\rangle\langle{\bf R}_{j}\rangle}{\sqrt{\left(\langle{\bf R}_{i}^{2}\rangle-\langle{\bf R}_{i}\rangle^{2}\right)\left(\langle{\bf R}_{j}^{2}\rangle-\langle{\bf R}_{j}\rangle^{2}\right)}}\label{CorrelCoefEq}\end{equation}
where ${\bf R}_{i}$ is the vector of the time series of log-returns,
$R_{i}(t)=\ln P_{i}(t)-\ln P_{i}(t-1)$ the log-return and $P_{i}(t)$
the daily closure price of stock $i$ at day $t$. The notation $\langle\cdots\rangle$
means an average over time $\frac{1}{T}\sum_{t'=t}^{t+T-1}\cdots$,
where $t$ is the first day and $T$ is the length of our time series.

This coefficient can vary between $-1\leq\rho_{ij}\leq1$, where $-1$
means completely anti-correlated stocks and $+1$ completely correlated
stocks. If $\rho_{ij}=0$ the stocks $i$ and $j$ are uncorrelated.
The coefficients form a symmetric $N\times N$ matrix with diagonal
elements equal to unity.

Following Onnela {\it et al.} \cite{Onnela_PRE68_2003,Onnela_EPJB285_2002},
we analyse the distribution of correlations in time. The first moment
is the mean correlation:

\begin{equation}
\overline{\rho}=\frac{2}{N(N-1)}\sum_{i<j}\rho_{ij}\label{MeanCorrel}\end{equation}
Other moments are similarly defined, the variance:

\begin{equation}
\lambda_{2}=\frac{2}{N(N-1)}\sum_{i<j}(\rho_{ij}-\overline{\rho})^{2},\label{VarianceCorrel}\end{equation}
the skewness:

\begin{equation}
\lambda_{3}=\frac{2}{N(N-1)\lambda_{2}^{3/2}}\sum_{i<j}(\rho_{ij}-\overline{\rho})^{3},\label{SkewnessCorrel}\end{equation}
and the kurtosis:

\begin{equation}
\lambda_{4}=\frac{2}{N(N-1)\lambda_{2}^{2}}\sum_{i<j}(\rho_{ij}-\overline{\rho})^{4}.\label{KurtosisCorrel}\end{equation}
Evaluation of these moments for time windows of width $T$ reveals
the dynamics of the time series. The higher moments explain how the
variance of correlation coefficients increase or decrease and how the 
skewness and kurtosis of the distribution changes. As we
will see in Section \ref{Sect4}, these moments show different behaviour
after crashes or financial days with significant news.

\subsection{Distances}

The metric distance, introduced by Mantegna \cite{Mantegna_EPJB193_1999},
is determined from the Euclidean distance between vectors, $d_{ij}=|\tilde{{\bf R}}_{i}-\tilde{{\bf R}}_{j}|$.
Here, the vectors $\tilde{{\bf R}}_{i}$ are computed from ${\bf R_{i}}$
by subtracting the mean and dividing by the standard deviation:

\[
\tilde{{\bf R}}_{i}=\frac{{\bf R}_{i}-<{\bf R}_{i}>}{\sqrt{\langle{\bf R}_{i}^{2}\rangle-\langle{\bf R}_{i}\rangle^{2}}}\]
Using the definition of correlation coefficient (eq. \ref{CorrelCoefEq}),
$\rho_{ij}=\tilde{{\bf R}}_{i}\cdot\tilde{{\bf R}}_{j}$ and noting
that $|\tilde{{\bf R}}_{i}|=1$ it follows that:

\[
d_{ij}^{2}=|\tilde{{\bf R}}_{i}-\tilde{{\bf R}}_{j}|^{2}=|\tilde{{\bf R}}_{i}|^{2}+|\tilde{{\bf R}}_{j}|^{2}-2\tilde{{\bf R}}_{i}\cdot\tilde{{\bf R}}_{j}=2-2\rho_{ij}\]
This relates the distance of two stocks to their correlation coefficient:

\begin{equation}
d_{ij}=\sqrt{2(1-\rho_{ij})}\label{Distance}\end{equation}
This distance varies between $0\leq d_{ij}\leq2$ where small values
imply strong correlations between stocks.

Following the procedure of Mantegna \cite{Mantegna_EPJB193_1999},
this distance matrix is now used to construct a network with the essential
information of the market. This network is a minimal spanning tree
(MST) with $N-1$ links connecting $N$ nodes. The nodes represent
stocks and the links are chosen such that the sum of all distances
(normalized tree length) is minimal. We perform this computation using
Prim's algorithm \cite{Prim}. %An example of which is shown in figure \ref{MST}.

The normalized tree length, again following Onnela {\it et al.} \cite{Onnela_PRE68_2003,Onnela_EPJB285_2002},
is given by

\begin{equation}
L=\frac{1}{N-1}\sum_{d_{ij}\in\mathrm{\Theta}}d_{ij}\label{MeanLength}\end{equation}
 where $\Theta$ represents the MST. We also compute its higher moments
(variance, skewness and kurtosis) and compare with the equivalent
moments of the correlations.

\section{Determination of time parameters\label{Sect3}}

\begin{figure}[H]
\begin{center}
\epsfysize=80mm
\epsffile{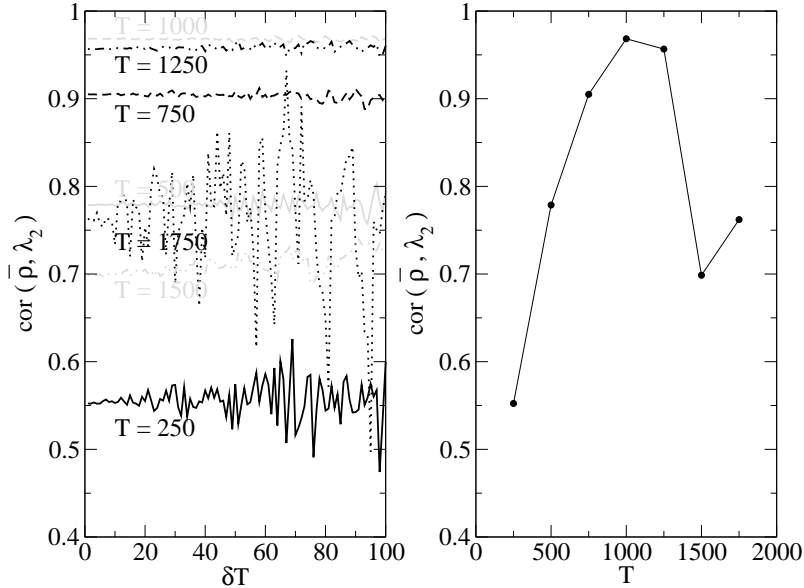}
\caption{Correlation between the first two moments
of the correlation coefficient (mean, eq. \ref{MeanCorrel} and variance,
eq. \ref{VarianceCorrel}) as a function of $T$ and $\delta T$.
The left graphic shows the correlation for different $T$ as function
of $\delta T$. The right graphic shows the correlation for $\delta T=1$,
as function of $T$.}
\label{Correl_deltaT_TT}
\end{center}
\end{figure}

Depending on the length of the time series, the correlation coefficient
between two stocks changes. Thus the distance between the two stocks
will be different and the MST constructed will have different characteristics.
In order to select appropriate values for the size of time windows
($T$) and window step length parameter ($\delta T$) we looked at
early studies in this field. As shown previously \cite{Onnela_PRE68_2003},
the first and second moment of the correlations (mean correlation
and variance) are strongly correlated. Taking this into account,
we computed the value of this correlation as a function of $T$ and
$\delta T$ (Figure \ref{Correl_deltaT_TT}).

Clearly, for all $T$, the correlation between the two moments is
not only positive but strong, above $0.9$ for $T=750$, $T=1000$
and $T=1250$. Apart from $T=250$ and $T=1750$ there are only very
small fluctuations for the correlation value, when we vary $\delta T$.
Since when we increase $\delta T$, we are essentially removing points
from our data, we decided to use the smallest value of $\delta T$
($1$ day) in all of the following. 

\begin{figure}[H]
\begin{center}
\epsfysize=80mm
\epsffile{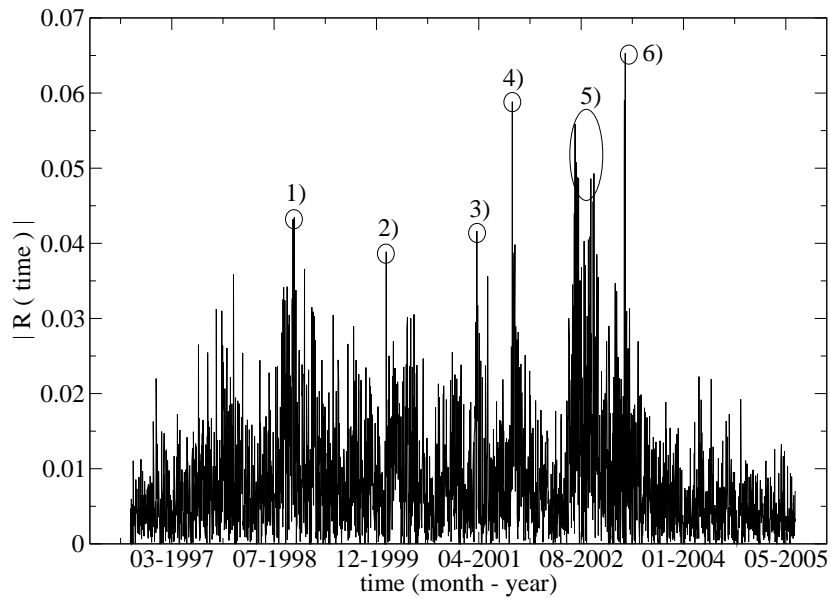}
\caption{Absolute return of the FTSE Index. Higher
values indicate special days like beginning of wars or crashes. $1)$
Russian crash; $2)$ NASDAQ crash; $3)$ Beginning of US recession;
$4)$ $11th$ September 2001; $5)$ Stock Market downturn of 2002;
$6)$ Beginning of Iraq War.}
\label{Return_FTSEIndex}
\end{center}
\end{figure}

Some events such as wars or crashes occurred during the period of study
and are noted in Figure \ref{Return_FTSEIndex} that shows the absolute
return of the FTSE Index. After these occurrences, which have a negative
effect on stock values, all the stocks seems to follow each other, and
both the correlation between them and mean correlation increase \cite{Onnela_PhysA247_2003}.
Now even if the correlation between mean correlation and variance increases
when we increase $T$, the curve of mean correlation is based on less information.
So the choice of $T$ becomes something of a compromise. We choose $T=500$
(2 years) and $\delta T=1$ (1 day) to compute our moments for the
correlations and distances.

\section{Analysis of Global Portfolio of FTSE100 index\label{Sect4}}

\begin{figure}[H]
\begin{center}
\epsfysize=60mm
\epsffile{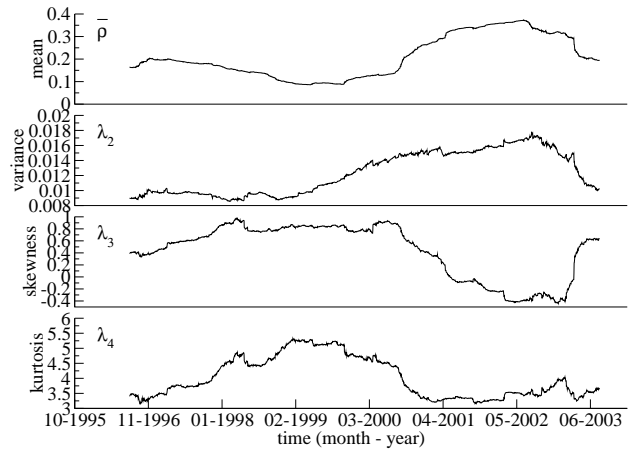}
\caption{Mean (eq. \ref{MeanCorrel}), variance (eq.
\ref{VarianceCorrel}), skewness (eq. \ref{SkewnessCorrel}) and kurtosis
(eq. \ref{KurtosisCorrel}) of the correlation coefficients. We use
time windows of length $T=500$ days and window step length parameter
$\delta T=1$ day. For each moment ($(T_{total}-T)/\delta T=$)$1822$
data points are shown.}
\label{Correl_Moments}
\end{center}
\end{figure}

The time dependence of the mean correlation, the normalized tree length
and the higher moments associated with these two quantities were studied
for a time window, $T=500$ and window step length, $\delta T=1$. Figure
\ref{Correl_Moments} shows that the mean and variance of the correlation
coefficients are highly correlated ($0.779$), the skewness and kurtosis
are also highly correlated and the mean and skewness are anti-correlated.
This implies that when the mean correlation increases, usually after
some negative event in market, the variance increases.
Thus the dispersion of values of the correlation
coefficient is higher. The skewness is almost always positive, which
means that the distribution is asymmetric, but after a negative event
the skewness decreases, and the distribution of the correlation 
coefficients becomes more symmetric.

\begin{figure}[H]
\begin{center}
\epsfysize=60mm
\epsffile{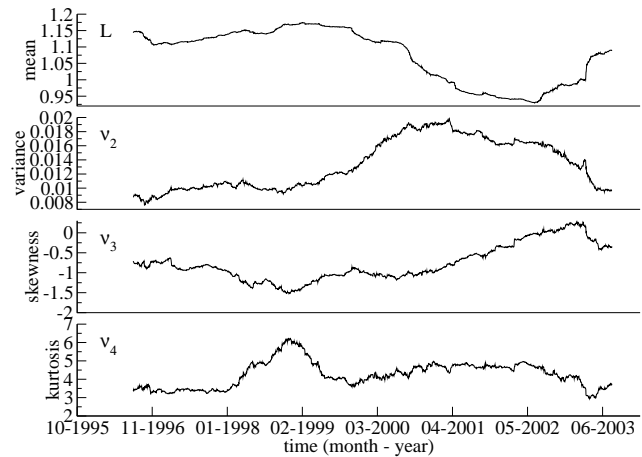}
\caption{Mean (eq. \ref{MeanLength}), variance, skewness
and kurtosis of the normalized tree length. We use time windows of
length $T=500$ days and window step length parameter $\delta T=1$
day. For each moment there are $1822$ points represented.}
\label{Length_Moments}
\end{center}
\end{figure}

From Figure \ref{Length_Moments}, we see how the normalized tree length
changes with time. As expected from equation \ref{Distance}, when
the mean correlation increases, the normalized tree length decreases
and vice versa. Here, the mean and the variance of the normalized
length of the tree are anti-correlated but the skewness and the mean
continue to be anti-correlated. This means that after some negative
event impacts the market, the tree shrinks, so the mean distance decreases
\cite{Onnela_PhysA247_2003}, the variance increases implying a higher
dispersion of the values of distance and the skewness, that is almost
always negative, increases showing that the distribution of the distances
of the MST gets more symmetric.

\begin{figure}[H]
\begin{center}
\epsfysize=70mm
\epsffile{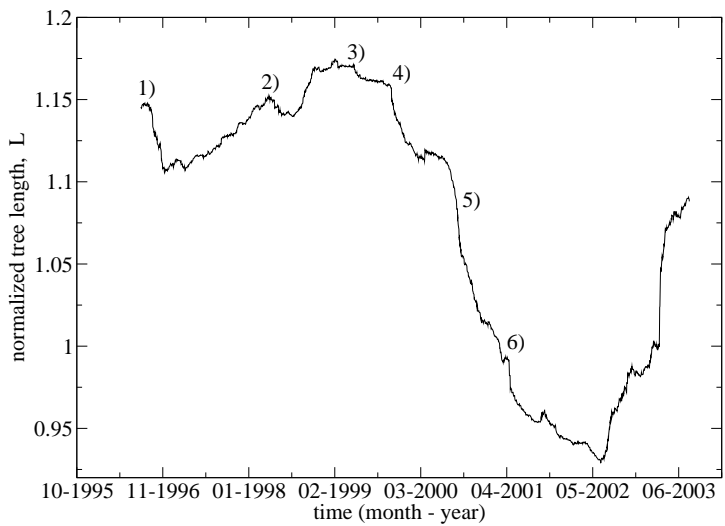}
\caption{Normalize tree length in function
of time. Different external events affect the market. $1)$ Russian
crash; $2)$ NASDAQ crash; $3)$ US recession; $4)$ $11th$ September
2001; $5)$ Stock Market Downturn of 2002; $6)$ Iraq War.}
\label{External_Events_in_time}
\end{center}
\end{figure}

Figure \ref{External_Events_in_time} is an enlarged version of the top graphic of Figure
\ref{Length_Moments}. This shows the mean of the normalized tree
length. As can be seen, after some of the events shown in Figure \ref{Return_FTSEIndex}:
the Russian Crash (October 1998), Dot-Com Bubble (March 2000), the
beginning of US recession (March 2001), attack to the Twin Towers
($11th$ September 2001), the Stock Market Downturn of 2002 with accounting
scandals (a long period between March 2002 and October 2002) and the
beginning of Iraq War (March 2003) the normalized tree length decreases.

\section{Sector Analysis\label{Sect5}}

A study of stocks such as the one considered here give us a insight into
the behavior with time of the market. A specific study of each
sector of the market is also of interest. We
have studied two different classifications. First we consider the old
classification for the London FTSE100, the FTSE Global Classification System
\cite{FTSEClassification}, that was in use from 2003 until the end
of 2005. This classification groups the stocks into $102$ Subsectors,
$36$ Sectors and $10$ Economic Groups. Our portfolio
is composed of $9$ economic groups and $27$ sectors: Resources (Mining,
Oil \& Gas), Basic Industries (Chemicals, Construction \& Building
Materials), General Industrials (Aerospace \& Defense), Non-cyclical
Consumer Goods (Beverages, Food Producers \& Processors, Health, Personal
Care \& Household Products, Pharmaceuticals \& Biotechnology, Tobacco),
Cyclical Services (General Retailers, Leisure \& Hotels, Media \&
Entertainment, Support Services, Transport), Non-cyclical Services
(Food \& Drug Retailers, Telecommunication Services), Utilities (Electricity,
Utilities-Others), Financials (Banks, Insurance, Life Assurance, Investment
Companies, Real Estate, Speciality \& Other Finance) and Information
Technology (Software \& Computer Services). 

The second classification studied is the new classification adopted
by FTSE since the beginning of 2006, the Industry Classification Benchmark
\cite{ICBClassification} created by Dow Jones Indexes and FTSE. This
classification is divided into $10$ Industries, $18$ Supersectors,
$39$ Sectors and $104$ Subsectors. Our portfolio
is composed of $10$ industries and $28$ sectors: Oil \& Gas (Oil
\& Gas Producers), Basic Materials (Chemicals, Mining), Industrials
(Construction \& Materials, Aerospace \& Defense, General Industrials,
Industrial Transportation, Support Services), Consumer Goods (Beverages,
Food Producers, Household Goods, Tobacco), Health Care (Health Care
Equipment \& Services, Pharmaceuticals \& Biotechnology), Consumer
Services (Food \& Drug Retailers, General Retailers, Media, Travel
\& Leisure), Telecommunications (Fixed Line Telecommunications, Mobile
Telecommunications), Utilities (Electricity, Gas Water \& Multiutilities),
Financials (Banks, Nonlife Insurance, Life Insurance, Real Estate,
General Financial, Equity Investment Instruments, Nonequity Investment
Instruments) and Technology (Software \& Computer Services).

For the old classification, the four economic groups with more stocks
are the Non-cyclical Consumer Goods ($13$), Cyclical Services ($21$),
Non-cyclical Services ($6$) and Financials ($13$). For each one
of these groups we have repeated the above analysis of moments.

\begin{figure}[H]
\begin{center}
\epsfysize=80mm
\epsffile{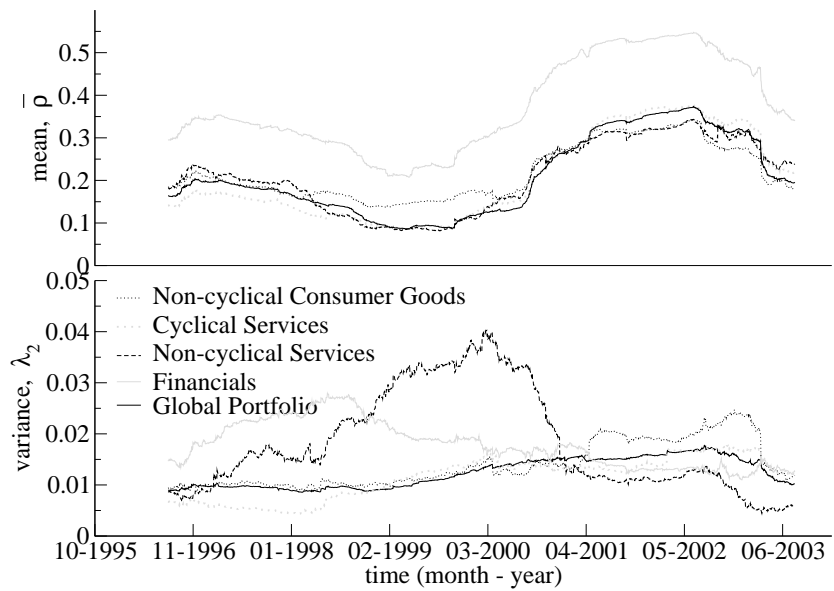}
\caption{Mean and variance of the correlation
coefficients for different economic groups, from the FTSE Global Classification
System, in comparison with the global portfolio.}
\label{Sector_Moments_Correl}
\end{center}
\end{figure}

As can be seen, not all the economic groups behave like the global
portfolio. Looking at the mean correlation, the Financial group is
much more correlated than all the other groups. If we analyse the
variance, the Financial and Non-cyclical Services groups loose the
global property where the first two moments of the correlation coefficients
are correlated.

\begin{figure}[H]
\begin{center}
\epsfysize=80mm
\epsffile{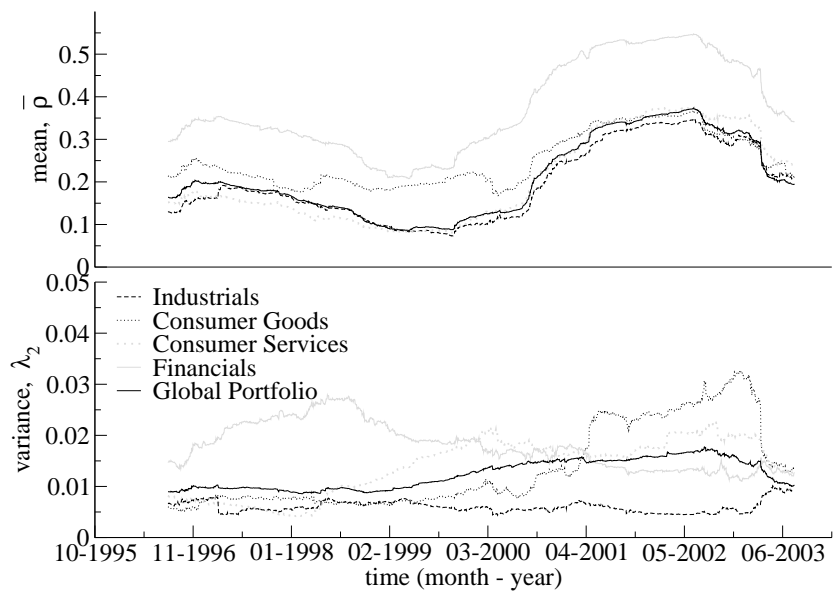}
\caption{Mean and variance of the correlation
coefficients for different industries, from the ICB, in comparison
with the global portfolio.}
\label{Sector_Moments_Correl_ICB}
\end{center}
\end{figure}

For the new classification, the four industries with more stocks are
the Industrials ($10$), Consumer Goods ($9$), Consumer Services
($18$) and Financials ($13$). The mean and variance of the correlation
coefficients for these industries are presented in Figure \ref{Sector_Moments_Correl_ICB}. 

With this classification, all the industries loose the global property
where the first two moments of the correlation coefficients are correlated.

\section{Minimal Spanning Trees\label{Sect6}}

For a topological view of the market we plot the MST with all the
nodes (stocks) and links between them (distances). For each classification
we analyse the cluster formation of different economic groups (FTSE
Global Classification System) or industries (ICB).

Starting with the analysis due to the old classification we represent
each economic group by a different symbol: Resources ($\blacksquare$),
Basic Industries ($\vartriangle$), General Industrials ($\blacklozenge$),
Non-cyclical Consumer Goods ($\square$), Cyclical Services ($\blacktriangle$),
Non-cyclical Services ($\lozenge$), Utilities ($\bullet$), Financials
(gray $\circ$) and Information Technology ($\circ$).

\begin{figure}[H]
\begin{center}
\epsfysize=80mm
\epsffile{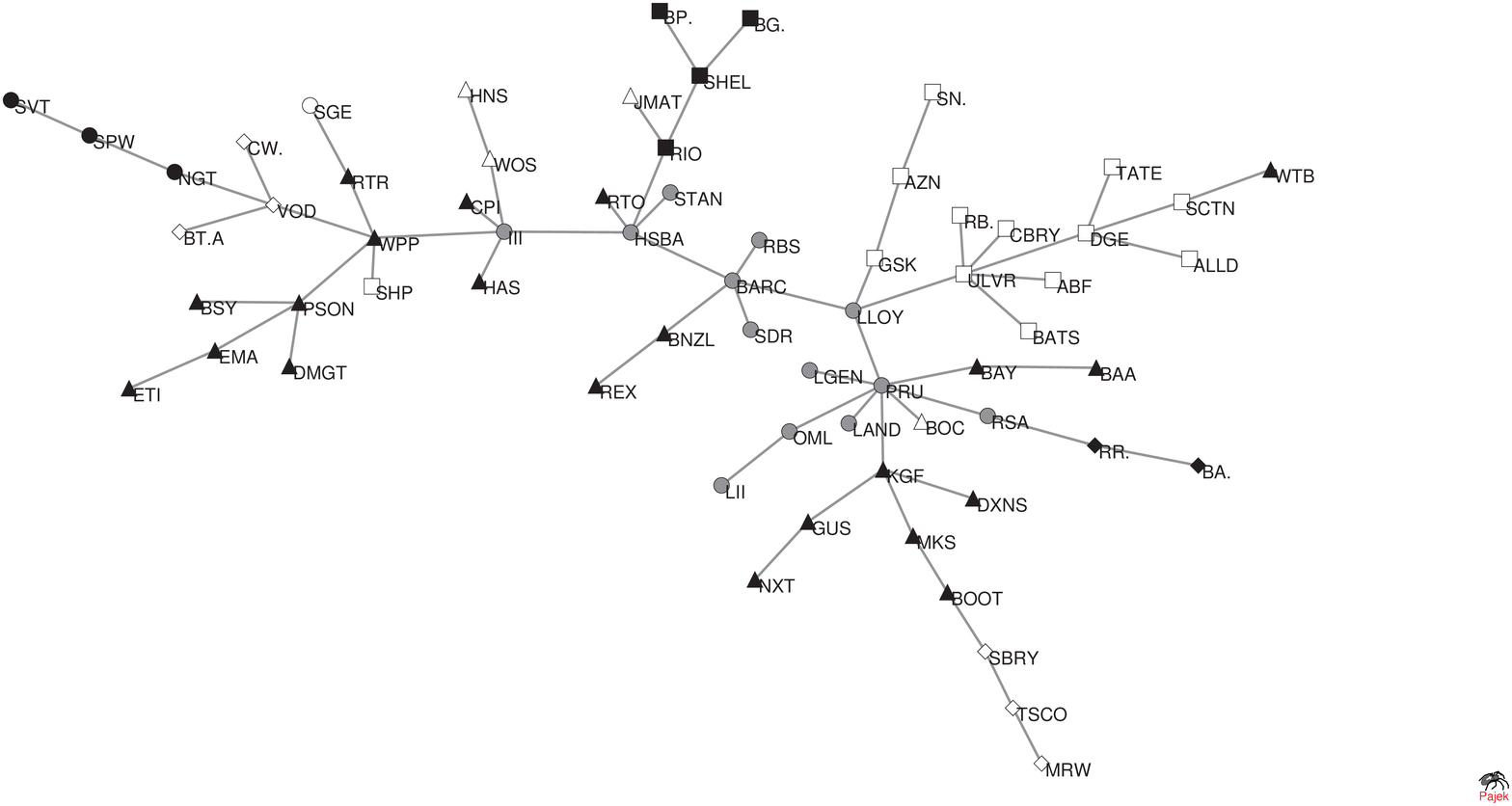}
\caption{Minimal Spanning Tree for the FTSE100 stocks. The length
of the time series used to compute this tree is 2322 days. Each symbol
correspond to a specific economic group from the FTSE Global Classification
System.}
\label{MST}
\end{center}
\end{figure}

Figure \ref{MST}, shows the MST with clusters of specific economic
groups. Stocks from the Financial group are the backbone of this tree.
It seems that all the other groups are connected to this one. The
Financials, Resources, Utilities and General Industrials groups have
all their stocks connected together. However for other groups divisions
of stocks in sectors are apparent. For example, in the Non-cyclical
Services, the Food \& Drug Retailers are completely separated from
the Telecommunication Services. Within Cyclical Services, the General
Retailers, Media \& Entertainment and Transports are $3$ different
clusters and the Support Services are isolated stocks connected to
the Financial branch. In Non-cyclical Consumer Goods, the Health and
Pharmaceuticals \& Biotechnology form one cluster whereas Beverages,
Tobacco, Food Producers \& Processors and Personal Care \& Household
Products form another. 

\begin{figure}[H]
\begin{center}
\epsfysize=80mm
\epsffile{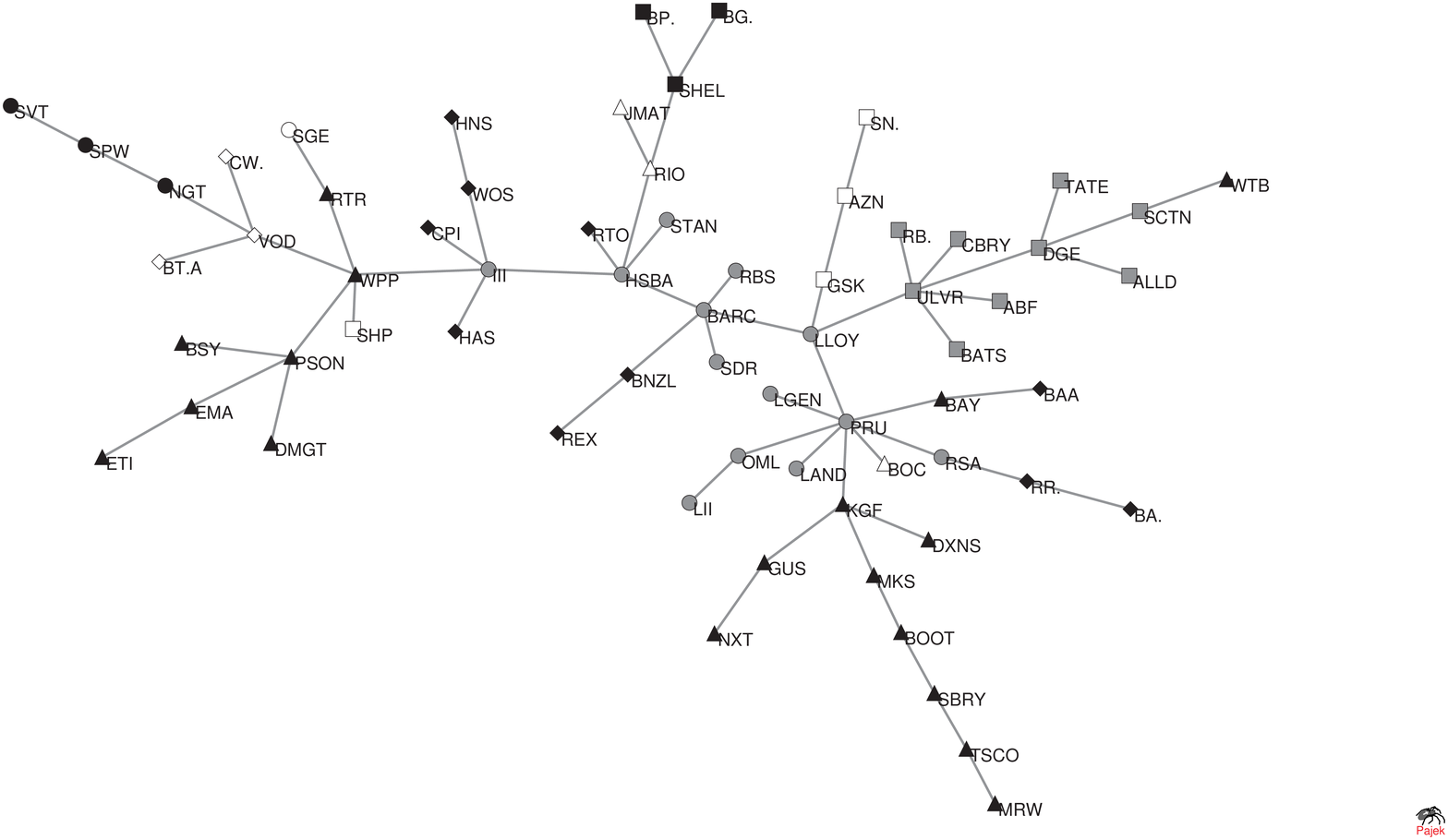}
\caption{Minimal Spanning Tree for the FTSE100 stocks. The
length of the time series used to compute this tree is 2322 days.
Each symbol correspond to a specific industry from the ICB.}
\label{MST_ICB}
\end{center}
\end{figure}

For the new classification, we represent each industry by a symbol:
Oil \& Gas ($\blacksquare$), Basic Materials ($\vartriangle$), Industrials
($\blacklozenge$), Consumer Goods (gray $\square$), Health Care ($\square$),
Consumer Services ($\blacktriangle$), Telecommunications ($\lozenge$),
Utilities ($\bullet$), Financials (gray $\circ$) and Technology
($\circ$). The MST is represented in Figure \ref{MST_ICB}. The Financial
industry has the same stocks as the one in the old classification,
so it still works as the backbone of the tree. Financials, Oil \&
Gas, Utilities, Telecommunications and Consumer Goods have all their
stocks connected together. In the Consumer Services, the supersectors
Retail and Media are two big clusters but they are not connected together.
The other supersector from this industry, the Travel \& Leisure is
disperse in the tree. Health Care industry is almost one cluster,
but the stock SHP is not connected to the others. In the Industrials
industry the stocks from the Support Services sector are always connected
to the Financial industry. The other stocks in this sector are located in 
isolation at other points within the tree.

The new classification adopted by FTSE in January 2006 clearly mimics much
more closely the MST results as we can see from Figures \ref{MST} and \ref{MST_ICB}. 
The implementation of the new supersector groups ensures that apart from
some notable exceptions stocks from the same supersector are now connected.
It is possible that the few stocks separated from their main cluster are
isolated by chance and over time they will join the appropriate clusters.
However there could be other more fundamental reasons for their separation.
Further study of both the dynamics of the MST correlations together with
their economic indices ({\it e.g.} PE ratio, earnings, etc) that characterize the
businesses concerned is necessary to resolve these issues. 
Nevertheless it seems clear from this analysis that the MST approach is one
that should complement current approaches to the development of stock
taxonomy.

Coronnello {\it et al.} \cite{Mantegna_Acta_Polonica_2005} have studied
the topology of the London FTSE using daily and intra-day data for
$N=92$ stocks, from year 2002. The MST for daily data looks quite
different from the one shown in Figure \ref{MST}. Using our data
and studying the MST for each year, we can see that for 2002, the
main hubs of the MST are BARC, RBS and SHEL, each of them with $11$,
$8$ and $7$ links, respectively. The simple inclusion of BARC in
our study (not included in the portfolio of \cite{Mantegna_Acta_Polonica_2005})
gives a quite different network. But the main clusters are the same
in the two studies.

\section{Numerical Simulations of MST\label{Sect7}}

\begin{figure}[H]
\begin{center}
\epsfysize=80mm
\epsffile{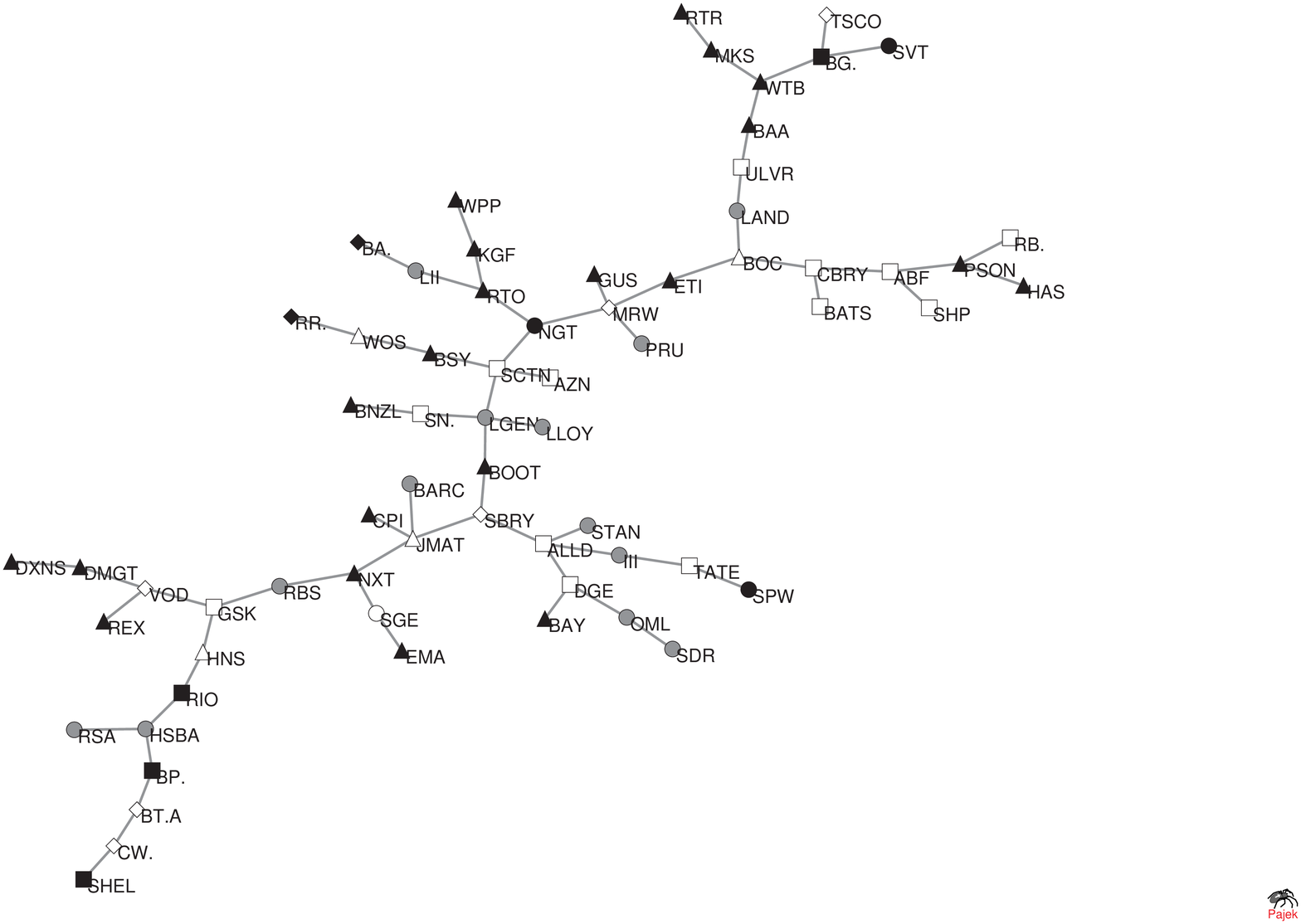}
\caption{Minimal Spanning Tree for $67$ random time series
using random variables from a Gaussian distribution.}
\label{MST_Random}
\end{center}
\end{figure}

In order to examine further the underlying nature of the time series we use now random
time series computed from two different models. Modeling the log-returns
as random numbers from a specific distribution, we can compute the
correlations, distances and trees for this random series. As in \cite{Mantegna_PRE68_046130,Mantegna_EPJB363_2004},
our first approach was to consider the returns as random variables derived from
a Gaussian distribution. So, using the real mean value, $\mu_{i}$ of each real time
series and the specific real variance, $\sigma_{i}$ we compute random series for our
{\it random market}:

\[
r_{i}(t)=\mu_{i}+\epsilon_{i}(t)\]
where $\epsilon_{i}(t)$ is the stochastic
variable from a Gaussian distribution with variance $\sigma_{i}$.
The MST for this random time series is represented in Figure \ref{MST_Random}.

This MST shows no clustering, the stocks are distributed randomly
in the network and there is no stock with more than $4$ links. To
create random time series with more real characteristics
we introduce a control term (the return of FTSE Index) and we compute
a one-factor model \cite{Mantegna_PRE68_046130,Mantegna_EPJB363_2004}:

\[
r_{i}(t)=\alpha_{i}+\beta_{i}R_{m}(t)+\epsilon_{i}(t)\]
where $\alpha_{i}$ and $\beta_{i}$ are parameters estimated by the least
square method from our data, $R_{m}(t)$ is the market factor (return
of FTSE Index) and $\epsilon_{i}(t)$ is the stochastic variable from
a Gaussian distribution with variance $\sigma_{i}$. The two factors
are calculated as:

\[
\alpha_{i}=\langle R_{i}(t)\rangle-\beta_{i}\langle R_{M}(t)\rangle\]

\[
\beta_{i}=\frac{cov(R_{i}(t),R_{M}(t))}{\sigma_{R_{M}}^{2}}\]
where $cov(\ldots,\ldots)$ is the covariance, $\sigma_{R_{M}}^{2}$
is the variance of the returns of FTSE Index and $R_{i}(t)$ is the
returns of real stock $i$.

\begin{figure}[H]
\begin{center}
\epsfysize=80mm
\epsffile{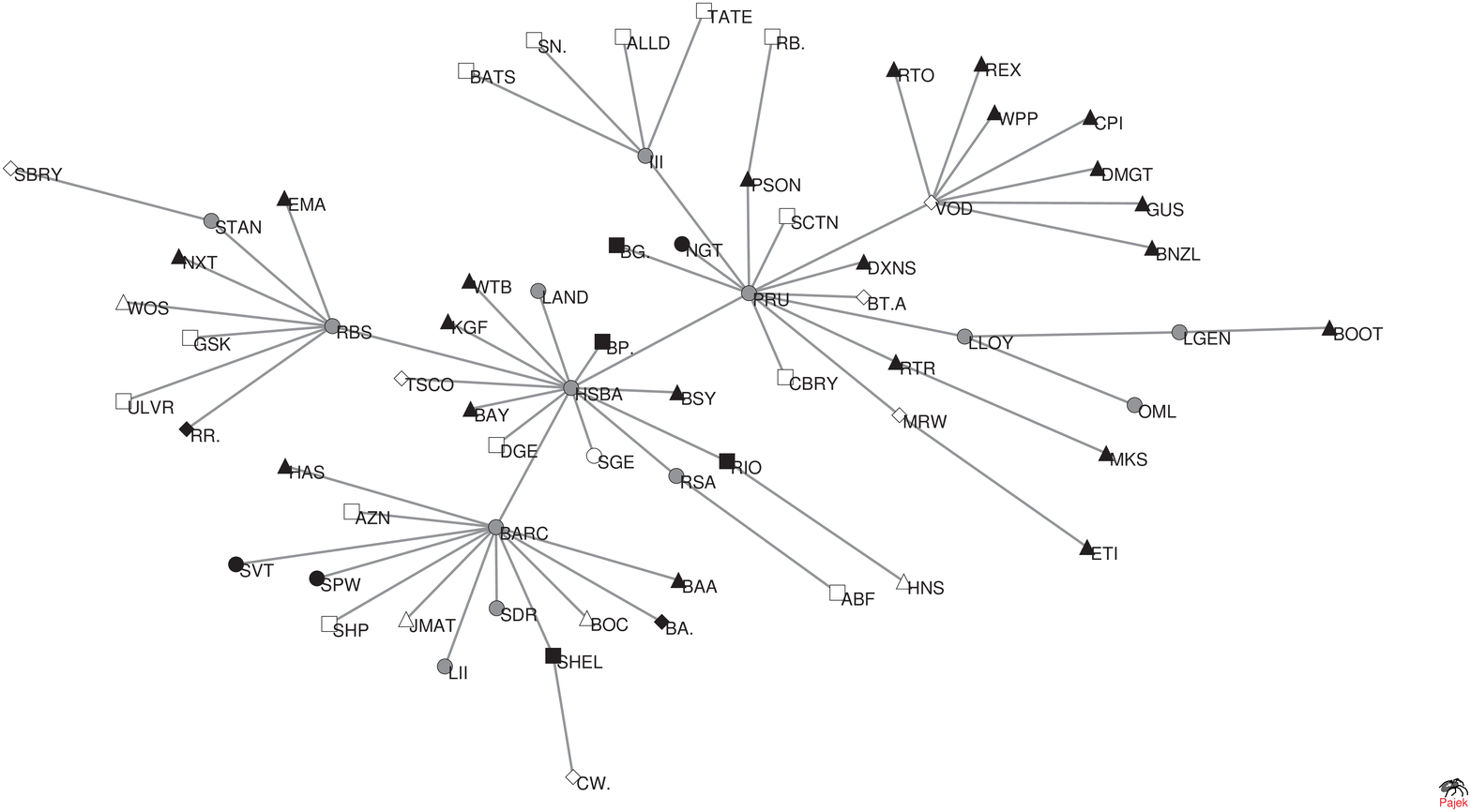}
\caption{Minimal Spanning Tree for $67$ random time
series using the one-factor model.}
\label{MST_Random_MM}
\end{center}
\end{figure}

The MST for random time series created using this model is shown
in Figure \ref{MST_Random_MM}.

This network is completely different from the previous random network.
Now we see that the stocks from the Financial group (gray $\circ$),
are all linked together. As in the MST for real data (figures \ref{MST} and \ref{MST_ICB})
they act as the backbone of the network.
However, the presence of $6$ nodes with up to $13$ links differs from
the topology of real data. A model to describe the time
series of log-return seems to lie somewhere between a completely random model
and an one-factor model. The completely random model does not give
much information. However, apart from producing the Financial group
backbone, the one-factor model shows similar topology with the MST
shown by Coronnello {\it et al.} \cite{Mantegna_Acta_Polonica_2005}
for intra-day data. This suggests that the network formed by intra-day
data are not fully formed, information as not yet sufficiently spread 
and correlations are not yet developed as they are found in
the networks formed using daily data.

\section{Conclusions\label{Sect8}}

In summary, we have studied the correlations between time series of
log-return of stocks from a FTSE100 portfolio and examine how these
change with both the size of the time series and time. The mean correlation
increases after external crises, and different moments feature 
correlations or anti-correlations as a result. For the study of specific
stocks of each sector we conclude that some sectors have different
feedback to the external events.

From the MST we can see that some stocks from the same sector cluster
together. This does not happen with all stocks from specific economic
groups or industries. It would seem from the MST analysis that the new FTSE classification
introduced in January 2006 offers a more logical clustering of the
different stocks as opposed to the previous classification scheme. However
from the MST it is clear that anomalies are still present that could affect
the building of optimum portfolios.

The structure of trees
generated from random time series differs significantly from real
markets. Furthermore there appears to be no obvious hub node. On the
other hand the one-factor model produces a MST where we can see hubs
with many links. This kind of structure is close to that obtained
using intra-day data. In future papers we shall assess changes in
the tree structures using one-factor model Levy distributions. We
shall also look at this issue by deriving analytic expressions linking
the moments of correlations to the moments of the lengths.

\begin{ack}
This publication has emanated from research conducted with the financial
support of Science Foundation Ireland (04/BRG/PO251). The authors also
acknowledge the help of COST (European Cooperation in the Field of
Scientific and Technical research) Action P10.
\end{ack}

\end{document}